\def\setup{\count90=0 \count80=0 \count91=0 \count92=0 \count85=0
\countdef\refno=80 \countdef\secno=85 \countdef\equno=90 \countdef\equnoA=92
\countdef\ceistno=91 }
\def\autoeq{ {\global\advance\count90 by 1} \eqno(\the\count90) }
\def\autoeqA#1{ {\global\advance\count92 by 1} \eqno(#1.\the\count92) }
\def\autoref{ {\global\advance\refno by 1} \kern -5pt [\the\refno]\kern 2pt}
\def\Gu{\Gamma_0(2)}
\def\e{\hbox{e}}
\def\circum#1{{ \kern -3.5pt $\hat{\hbox{#1}}$ \kern -3.5pt}}
\def\vline{{\vrule height8pt depth4pt}\; }
\def\Vline{{\vrule height13pt depth8pt}\; }
\def\sub#1{{\lower 8pt \hbox{$#1$}}}
\magnification=1200
\setup
\overfullrule=0pt
\centerline{  }
\rightline{DIAS-STP 98-10}
\rightline{cond-mat/9809294}
\rightline{10th February 1999}
\rightline{Revised Version}
%\rightline{Revision no. 1} 
\vskip 2cm
\centerline{\bf Modular Invariance, Universality and Crossover in the Quantum Hall Effect}
\vskip 1.2cm
\centerline{Brian P. Dolan}
\vskip .5cm
\centerline{\it Department of Mathematical Physics}
\centerline{\it National University of Ireland, Maynooth, Ireland}
\centerline{and}
\centerline{\it Dublin Institute for Advanced 
Studies}
\centerline{\it 10, Burlington Rd., Dublin, Ireland}
\vskip .5cm
\centerline{e-mail: bdolan@thphys.may.ie}
\vskip 1.5cm
%\baselineskip=1.5 \baselineskip
\centerline{\bf ABSTRACT}
\bigskip
\noindent An analytic form for the conductivity tensor in crossover between
two quantum Hall plateaux is derived, which appears to be  in good agreement with
existing experimental data. The derivation relies on an assumed
symmetry between quantum Hall states, a generalisation of the law
of corresponding states from rational filling factors to complex conductivity,
which has a mathematical expression in terms of an action of the modular
group on the upper-half complex conductivity plane. This symmetry implies
universality in quantum Hall crossovers. The assumption that the $\beta$-function
for the complex conductivity is a complex analytic function,
together with some experimental constraints, results in an
analytic expression for the crossover, as a function of the external magnetic
field.
\vfill\eject
%\baselineskip=1.5 \baselineskip
{\bf \S 1 Introduction}
\bigskip
The significance of symmetry groups in the understanding of physics, particularly
quantum physics, has steadily increased ever since their importance was first
realised. For example the representation theory of the rotation group in three
dimensions is immensely powerful in understanding the structure
of the periodic table of the elements, even before any underlying dynamics is studied, and the importance
of symmetry principles in elementary particle physics cannot be over emphasised.
It has been suspected for some time now,
\autoref\newcount\LutkenRossa\LutkenRossa=\refno
\autoref\newcount\LutkenRossb\LutkenRossb=\refno
\autoref\newcount\WilcekZee\WilcekZee=\refno
\autoref\newcount\GeorglinWallet\GeorglinWallet=\refno,
that a certain infinite discrete group, the modular group $\Gamma(1)$, 
acting on the upper-half complex conductivity plane,
$\sigma=\sigma_{xy}+i\sigma_{xx}$, plays an important r\circum{o}le
in understanding the phase structure of quantum Hall (QH) states. In particular
the fact that only odd denominator filling factors, $\nu=p/q$ where $q$
is an odd integer, are observed as stable quantum Hall states (for which
$\sigma_{xx}$ vanishes and $\sigma_{xy}=p/q$) can be interpreted as being due
a symmetry between states under the action of a particular sub-group of the
full modular group, often denoted by $\Gamma_0(2)$ in the literature,
[\the\LutkenRossb]\autoref\newcount\Lutken\Lutken=\refno.
It is the purpose of this paper to explore some of the consequences
of this observation for the crossover from one QH state to another.
\smallskip
If one assumes that the renormalisation group (RG) flow in the $\sigma$ plane,
induced by changing the strength of the external magnetic field, commutes with
the action of $\Gamma_0(2)$, one is led to conclude that fixed points
of $\Gamma_0(2)$ must be critical points of the RG flow, [\the\LutkenRossb],
[\the\Lutken]. This leads to many predictions concerning the phase structure,
including the selection rule for QH transitions $p_1q_2-p_2q_1=\pm 1$, 
between filling factors $\nu_1=p_1/q_1$ and $\nu_2=p_1/q_2$
\autoref\newcount\BD\BD=\refno. In [\the\LutkenRossb], [\the\Lutken]
and [\the\BD] only the critical points and phase structure were discussed,
nothing was said about the form of the crossover between QH sates beyond
the observation that $\Gamma_0(2)$ leads naturally to the semi-circle law
of \autoref\newcount\Ruzin\Ruzin=\refno and follows from the law
of corresponding states in \autoref\newcount\KLZ\KLZ=\refno.
Some suggestions about the functional form of the crossover
were made in
\autoref\newcount\LutkenBurgess\LutkenBurgess=\refno. 
\smallskip
In this work some assumptions will be made about the analytic
structure of the $\beta$-function for crossover between the  QH state 
with $\nu =1$ and the insulating phase with $\nu =0$. Using this as a template
one can then construct the $\beta$-function for crossover for any 
allowed QH transition, by acting with the appropriate element of $\Gamma_0(2)$
which maps $\nu=0$ and $\nu=1$ onto $\nu=p_1/q_1$ and $\nu=p_2/q_2$ 
respectively, with $p_2q_1-p_1q_2=1$. The validity of the assumptions can then be tested
by comparing the resulting predictions with existing experiments,
and the results are promising.
\smallskip
The most important assumption, as described in \S 3, is that
that the $\beta$-function is a complex analytic function of $\sigma$.
Together with symmetry under the action of $\Gu$  this restricts the
structure of the $\beta$-function to a very special form and some further,
rather mild, assumptions restrict it to an essentially unique function.
The assumption of analyticity of the $\beta$-function in the apparently
unrelated theory of Yang-Mills gauge theory was investigated in
\autoref\newcount\Ritz\Ritz=\refno,\autoref\newcount\LutkenLatorre\LutkenLatorre=\refno,
and much of the analysis presented here is based on ideas in these references.
The analytic form of the resulting crossover is shown in figure 3 for
the transition from $\nu=1$ to the insulating phase, and in figure 4 for
the crossover between $\nu=2$ and $\nu=1$, at various temperatures (the conductivity
is expressed in multiples of $e^2/h$).
The similarities between the theoretical predictions shown in
figures 3 and 4 here and the experimental results, shown in figures 1 and 2b of
\autoref\newcount\Shaharetal\Shaharetal=\refno are quite striking.
\smallskip
For ease of reference we give here the analytic form of the function
which is postulated to describe crossover between QH states $\nu=p_1/q_1$ and $\nu=p_2/q_2$,
which is the main result of this paper.
Let $\Delta\nu=\nu-\nu_c$, with $\nu={\bar n e\over B}$ the Landau level 
filling factor ($\bar n$ is the electron density and $B$ the external magnetic 
field) and $\nu_c={\bar ne\over B_c}$ the filling factor at the 
critical value of the magnetic field, $B_c$.
Define 
$\alpha:=(p_2-p_1)-(q_2-q_1)\nu_c$ and the linear function
$\zeta(\Delta\nu):=\alpha\{(q_1-q_2)\Delta\nu+\alpha\}$.
Then
$$\sigma(\Delta\nu)={p_2q_2\{ K^\prime(w)\}^2+ p_1q_1\{K(w)\}^2
+iK^\prime(w)K(w) \over \bigl[q_1^2\{K(w)\}^2+ q_2^2\{K^\prime(w)\}^2\bigr] },
\autoeq$$
\newcount\Cross\Cross=\equno
where $K^\prime(w)$ and $K(w)$ are complete elliptic integrals of
the second kind --- with modulus $w$ a
function of $\Delta\nu$ given by
$$w^2={1\over 2}\left\{1-\hbox{sign}(\Delta\nu)
\sqrt{1-\e^{-\bigl({A\Delta\nu\over\zeta(\Delta\nu) T^\mu}\bigr)^2}}\right\}.
\autoeq$$
\newcount\wdef\wdef=\equno
\noindent As usual the  complementary modulus $w^\prime$ is defined  by 
$(w^\prime)^2=1-w^2$ and $K^\prime(w)=K(w^\prime)$.
$T$ in equation (\the\wdef) denotes the temperature and $\mu$ the critical exponent described in 
\autoref\newcount\Pruiskena\Pruiskena=\refno, which is experimentally
determined to be $\mu=0.45\pm 0.05$, [\the\Shaharetal] (the definition is given in section 2).
$A$ is
a positive real constant, possibly  depending on other experimental parameters such as the
electron density and $B_c$. 
The arguments that lead to equation (\the\wdef) are strictly
only valid for small $\Delta\nu$, in principle there could
be corrections of order $(\Delta\nu)^4$ to the exponent, however
the range of $\Delta\nu$ required for crossover is small, $|\Delta\nu|<0.1$,
and equations (\the\Cross) and (\the\wdef) appear to give a good
fit to the experimental data as they stand --- the crossover that
they describe is shown graphically in figures 3 and 4.
Figure 3 is obtained by plotting the real and
imaginary parts of (\the\Cross) for the case $p_1=0$, $p_2=q_1=q_2=1$ and figure 4
for $p_1=q_1=q_2=1$, $p_2=2$, and these figures should be compared with the experimental
data in figures 1 and 2b of reference [\the\Shaharetal].
For the special case of the QH-insulator transition with $p_1=0$, and $q_1=q_2=p_2=1$
equation (\the\Cross) gives a resistivity,
$\rho=-1/\sigma$, whose longitudinal component is
$$\rho_{xx}(\Delta\nu)={K(w)\over K^\prime(w)},\autoeq $$
from which follows the relation
$$\rho_{xx}(\Delta\nu)={1\over \rho_{xx}(-\Delta\nu)},\autoeq$$
for this transition, since $w(-\Delta\nu)=w^\prime(\Delta\nu)$. 
This relation is well supported experimentally,
\autoref\newcount\STSCSS\STSCSS=\refno.
It seems remarkable that making the simplest possible assumption at every stage
leads to predictions which are so close to experiment and the bulk of this paper is 
devoted to the derivation of equation (\the\Cross).
\smallskip Explicitly, the assumptions which are used in the analysis are:
\bigskip
\item{i)} The law of corresponding states of [\the\KLZ] 
should be extended into the upper-half complex conductivity plane, 
$\sigma=\sigma_{xy}+i\sigma_{xx}$ (in units of $e^2/h$). This is
related to rotational invariance in the bulk and is
encoded mathematically into a group action on $\sigma$ --- specifically
the group is $\Gu$. 
\smallskip
\item{ii)} The action of $\Gu$ commutes with RG flow. This implies
universality of QH transitions.
\smallskip
\item{iii)} There are no extra critical points, other than those already
indicated by the experimental data and weak coupling expansions, i.e.
$\sigma=0,1,{1+i\over 2},i\infty$ and their images under the action of $\Gu$.
\smallskip
\item{iv)} In the limit of large $\sigma_{xx}$ (the weak coupling limit
of field theoretical models) $\sigma_{xy}$ does not run and the $\beta$-function
for $\sigma_{xx}$ is independent of $\sigma_{xy}$, finite and negative.
\smallskip
\item{v)} The $\beta$-functions for the RG flow that describes  crossover
between two QH plateaux, or between a QH state and the insulating phase,
are complex analytic functions when described in terms of a real
variable $s$ that is a monotonic real analytic function of the external
magnetic field. 
\bigskip
\noindent These assumptions alone actually lead to slightly more general form than equation
(\the\Cross), but demanding that Hall plateaux are approached as fast as
possible as the magnetic field is varied gives equation (\the\Cross).
\smallskip
In the next section the concept of universality and scaling in th QH effect is reviewed
and \S 3 contains a discussion of the r\circum{o}le of $\Gu$ in QH phenomena. Section
4 then introduces the assumption of analyticity into the discussion and it is
shown how this places strong restrictions on the form of the $\beta$-function.
Section 5 continues with a discussion of the RG behaviour near the critical
point in the crossover between Hall plateaux and it is shown that the $\beta$-function
is then essentially unique. In \S 6 the analytic form of the crossover is discussed
and equation (\the\Cross) is derived. Finally \S 7 gives a summary and outlook.
An appendix contains a review of some properties of the modular group, 
Jacobi $\vartheta$-functions and complete elliptic integrals which are relevant
to the discussion in the text but may be unfamiliar to a general audience.

\bigskip {\bf \S 2 Scaling and Universality in the Quantum Hall Effect}
\bigskip
The discussion of crossover relies heavily on the scaling analysis of [\the\Pruiskena],
building on ideas originally due to Khmel'nitskii
\autoref\newcount\Khem\Khem=\refno. This section is a review of the essential
points necessary for the later analysis. In [\the\Pruiskena] it was
suggested that there was a critical point in the crossover between two
QH states $\sigma=\nu_1$ and $\sigma=\nu_2$, near which the only relevant
variables are the external magnetic field, $B$, and the temperature, $T$.
The crossover is driven by varying $B$ at fixed $T$, and not by varying $T$ as
in the more familiar temperature driven transitions
(for a review of quantum phase transitions see
\autoref\newcount\QPT\QPT=\refno). The critical 
external field will be denoted by $B_c$. Thus the correlation length
diverges with critical exponent $\nu_\xi$,
$\xi\propto\vert B-B_c\vert^{-\nu_\xi}$ (the subscript $\xi$ on the exponent
is to avoid confusion with the Landau level filling factor, which
is also traditionally denoted by $\nu$ and appears frequently throughout
the text). Since $\sigma$ is dimensionless, in units
of $e^2/h$, it must be a function of a dimensionless ratio, $\tilde b=\Delta B/T^\mu$,
where $\Delta B=B-B_c$ and $\mu$ is a critical exponent which appears to be
universal in that it is the same for all transitions. The correlation length, $\xi$,
is related to $\tilde b$ through a characteristic length, $L$. 
For low magnetic field strengths $L$ can be taken to be the electron scattering 
length, [\the\Khem], which diverges as $T\rightarrow 0$ for an infinite size system.
The scaling form for
the way in which electron scattering length depends on the temperature is, [\the\Pruiskena],
$$L(T)\propto T^{-p/2},\autoeq$$
\newcount\Lscale\Lscale=\equno
where $p$ is the inelastic scattering length exponent.
For high field strengths a more appropriate characteristic length scale is the magnetic
length, $\sqrt{\hbar\over eB}$.
In any case $L$ would be expected to be a function of both $T$ and $B$ in general.
\smallskip
Now the ratio $l=L/\xi$ is dimensionless
and requiring that the correlation length is related to the magnetic field by
the usual scaling relation
$$\xi\propto\vert\Delta B\vert^{-\nu_\xi},\autoeq$$
\newcount\Bscale\Bscale=\equno
one has $\vert \tilde b\vert\propto l^{1/\nu_\xi}$. 
%Since it is the external field and not the temperature that drives the crossover
%in a quantum phase transition, the correlation length should be  independent of the
%temperature which requires the scaling relation
%$$p=2\mu\nu_\xi.\autoeq$$
%\newcount\slaw\slaw=\equno
Experimentally $\mu=0.45\pm0.05$, [\the\Shaharetal]
and $\nu_\xi=2.02$.
\smallskip
The analytic from of the crossover is then described by a function $\sigma(\tilde b)$ and
the $\beta$-functions for the longitudinal and transverse conductivities $\sigma_{xx}$
and $\sigma_{xy}$ can be  defined  as
$$\beta_{xx}(\sigma_{xx},\sigma_{xy})={d\sigma_{xx}\over d\tilde b} \qquad
\beta_{xy}(\sigma_{xx},\sigma_{xy})={d\sigma_{xy}\over d\tilde b}.\autoeq$$
\newcount\realbetas\realbetas=\count90
These can be combined into a single complex function $\beta=\beta_{xy}+i\beta_{xx}$
so that
$$\beta(\sigma,\bar\sigma)={d\sigma\over d\tilde b}.\autoeq$$
\newcount\complexbeta\complexbeta=\equno
Very little is known about the functional form of (\the\realbetas). A flow
diagram was suggested in [\the\Khem] and used in [\the\Pruiskena].
Asymptotic forms for $\sigma_{xx}$ and $\sigma_{xy}$ when $\sigma_{xx}$ is
large  have been calculated in
\autoref\newcount\Pruiskenb\Pruiskenb=\refno, based on an effective action which contains
a topological term. The interaction strength  is $1/\sigma_{xx}$, so that large $\sigma_{xx}$
is the weak coupling regime, and the topological term has coupling $\sigma_{xy}$, the
quantization of which explains
the quantization of the Hall states. In a perturbative calculation, $\sigma_{xy}$ is
constant and $\beta_{xx}$ depends only on $\sigma_{xx}$, but it is argued in [\the\Pruiskenb]
that non-perturbative instanton effects produce the more general form (\the\realbetas).
The same is true for supersymmetric Yang-Mills gauge theories, as revealed by the work 
of Seiberg and Witten
\autoref\newcount\SW\SW=\refno, on which [\the\Ritz] and [\the\LutkenLatorre] are based.
Reference [\the\SW] itself relied heavily on a conjecture that the modular
group should be a symmetry for supersymmetric Yang-Mills theory,
\autoref\newcount\MontonenOlive\MontonenOlive=\refno.
A non-perturbative form for $\beta(\sigma,\bar\sigma)$, based on $\Gu$ symmetry,
was suggested in [\the\LutkenBurgess].
\smallskip
In the article by Pruisken in [\the\Pruiskenb] the $\beta$-functions are calculated as derivatives
with respect to $l$ rather than $\tilde b$ --- this is quite legitimate and just
reflects the fact that there are
different ways to define $\beta$-functions. 
Another possibility
%\autoref\newcount\SGCS\SGCS=\refno 
is to use the Landau
level filling factor $\nu={\bar n e\over B}$, where $\bar n$ is the electron density.
Clearly 
$$\Delta\nu:=\nu-\nu_c=-{\bar n e\Delta B\over B_c(B_c+\Delta B)}
\approx -{\bar n e\Delta B\over (B_c)^2},\autoeq $$
where  $\nu_c={\bar n e\over B_c}$ is the critical filling factor,
and $\beta$-functions could be defined using $v:={\bar n e\Delta \nu\over T^\mu}$ as 
$$\beta(\sigma,\bar\sigma)={d\sigma\over dv}.\autoeq$$
\newcount\betav\betav=\equno
In the following we shall try to
be as general as possible and define
$$\beta(\sigma,\bar\sigma)={d\sigma\over ds},\autoeq$$
\newcount\nabeta\nabeta=\equno
where $s$ is a monotonic function of $\tilde b$ or $v$, to be determined.
\bigskip
{\bf \S 3 The Modular Group and the Quantum Hall Effect}
\bigskip
This section reviews the essential ingredients of the
action of the group $\Gu$ on the complex conductivity
plane ---  more details can be found in the appendix.
The full modular group, $\Gamma(1)$, was studied for possible significance in the QH effect in
[\the\LutkenRossa], inspired by observations originally due
to Cardy and Rabinovici\autoref\newcount\CardyRabinovici\CardyRabinovici=\refno
\newcount\Cardy\Cardy=\refno concerning the extension of Kramers-Wannier duality  to
a model with two parameters, one of which was topological in nature. Sub-groups
of the modular group, resulting in a more restrictive  symmetry, were analysed
in [\the\LutkenRossb] [\the\WilcekZee] [\the\GeorglinWallet] and [\the\Lutken], 
and in [\the\LutkenRossb]
and [\the\Lutken]
attention was focused on the particular group $\Gu$, which was further developed in
[\the\BD]. 
\smallskip
The basic assumption is that the infinite discrete group $\Gu$ has a natural action on the
upper-half complex conductivity plane, parameterised by $\sigma=\sigma_{xy}+i\sigma_{xx}$
in units of $e^2/h$, which is a symmetry of the partition function (a generalisation of
Kramers-Wannier duality) and commutes with the RG flow on the $\sigma$-plane.
The group $\Gu$ can be represented by $2\times2$ matrices of the from
$\pmatrix{a&b\cr 2c&d\cr}$ where $a,b,c$ and $d$ are integers with $ad-2bc=1$
(which requires $a$ and $d$ to be odd). For
$\gamma=\pmatrix{a&b\cr 2c&d\cr}\in\Gu$ the action is 
$\gamma:\sigma\rightarrow{a\sigma+b\over 2c\sigma+d}$. Thus $\gamma$ maps the QH state
with filling factor
$\nu=p/q$, which has $\Re(\sigma)=\nu$ and $\Im(\sigma)=0$, to $\nu={ap+bq\over 2cp+dq}$.
Note that $2cp+dq$ is odd if $q$ is, but this would not be true if the
factor of $2$ were omitted --- this is one way of seeing why it is the sub-group $\Gu$ that is
relevant to the QH effect rather than the full modular group, for which the $2$
would not be present.
\smallskip
All $\Gu$ transformations can be generated by repeated applications of the two
operators $U:\sigma\rightarrow\sigma+1$ and $V:\sigma\rightarrow{\sigma\over 2\sigma+1}$,
the former being a generalisation of the Landau level addition transformation
rule of [\the\KLZ], from real filling factors to the 
upper-half complex conductivity plane, and the latter being a similar generalisation of the 
flux attachment rule of the same reference. 
This generalisation is natural, and is indeed forced on us if the
law of corresponding states is combined with rotational
invariance in the bulk of a two dimensional sample. To see
this observe that the law of corresponding
states is formulated in terms of a discrete set of transformations on
quantum Hall states labelled by $\nu$, or $\sigma_{xy}$, 
$\sigma_{xy}\rightarrow\sigma_{xy}+1$ and 
$\sigma_{xy}\rightarrow{\sigma_{xy}\over 2\sigma_{xy}+1}$.
But $\sigma_{xy}$
is only one component of a tensor, 
$\sigma_{ij}=\left(\matrix{\sigma_{xx} & \sigma_{xy}\cr\sigma_{yx} & \sigma_{yy}\cr}\right)$. Two dimensional rotational invariance demands that $\sigma_{xx}=\sigma_{yy}$
and $\sigma_{xy}=-\sigma_{yx}$, where the latter is a parity violating
effect induced by the external field. It is incompatible with
rotational invariance to have a transformation acting on one component 
of a tensor, $\sigma_{xy}$, but not on the other,
$\sigma_{xx}$. This causes no problems on the Hall plateau, where 
$\sigma_{xx}=0$, but is inconsistent when $\sigma_{xx}\ne 0$. To see how to 
extend the law of corresponding sates to non-zero $\sigma_{xx}$, in is
convenient use complex co-ordinates in two dimensions,
$$z=x+iy, \qquad \bar z =x-iy,$$ in terms of which
$$\sigma_{ij}=\left(\matrix{\sigma_{xx} & \sigma_{xy}\cr
       -\sigma_{xy} & \sigma_{xx}\cr}\right) \rightarrow
\left(\matrix{0& \sigma_{xx}+i\sigma_{xy}\cr
       \sigma_{xx}-i\sigma_{xy} & 0\cr}\right)=
\left(\matrix{0&i\bar\sigma\cr -i\sigma&0\cr}\right).\autoeq$$
Thus, in complex co-ordinates, the tensor $\sigma_{ij}$ is described
by $\sigma=\sigma_{xy}+i\sigma_{xx}$ (with $\sigma_{xx}\ge 0$ 
and its complex conjugate.
The extension of the law of corresponding states to this complex conductivity
leads naturally to the group $\Gamma_0(2)$. It is noteworthy that the
the law of corresponding states, [\the\KLZ], was introduced in the quantum Hall
effect at almost exactly the same time as the group
$\Gamma_0(2)$, [\the\LutkenRossb], though from completely different
motivations --- both sets of authors appear to have been unaware
of each other's work.
\smallskip

Because the group action is assumed to
commute with the RG flow the critical exponents near a critical point $\sigma_c$ are
identical with those of $\gamma(\sigma_c)$, in accord with the universality
hypothesis of [\the\Pruiskena]. A flow diagram compatible with $\Gu$ is shown
in figure 1 --- such diagrams  were first produced
in [\the\LutkenRossb] and [\the\Lutken] and  their similarity to the diagrams
proposed in [\the\Pruiskena] and [\the\Khem], without any use of $\Gu$ but
using more specific  reasoning, is manifest. 
\smallskip
Under the action of $\Gu$ the $\beta$-function, (\the\nabeta), necessarily transforms as
$$\beta(\gamma(\sigma),\gamma(\bar\sigma))={d\over ds}\left({a\sigma+b\over 2c\sigma+d}\right)
={1\over (2c\sigma+d)^2}\beta(\sigma,\bar\sigma),\autoeq$$
\newcount\modbeta\modbeta=\equno
since $ad-2bc=1$.
It is thus very tempting to to assume that $\beta$ is a complex {\it analytic} function
of $\sigma$ on the upper-half complex plane as such functions, called modular functions
of weight $-2$, have been extensively studied in the mathematical literature 
and have many remarkable properties
\footnote*{More accurately a modular form is meromorphic in $q=\e^{i\pi\sigma}$ 
rather than $\sigma$ itself.}\autoref\newcount\Rankin\Rankin=\refno. A
justification of this assumption from first principles, using a microscopic Hamiltonian
with specific interactions, will not be attempted here as such a programme would
be very ambitious. Rather the consequences of the analyticity assumption will be
examined and it will be shown that it gives rise to crossovers that are remarkably
similar to the experimental data --- thus giving impetus to the more difficult
problem of finding a microscopic explanation of analyticity. We shall adapt
some of the ideas used in [\the\Ritz], in applying $\Gu$ symmetry to the RG flow
of supersymmetric Yang-Mills theory, based on the seminal work of Seiberg and Witten, [\the\SW]. 
Analyticity means that (\the\modbeta) reads
$$\beta(\gamma(\sigma))={d\over ds}\left({a\sigma+b\over 2c\sigma+d}\right)
={1\over (2c\sigma+d)^2}\beta(\sigma).\autoeq$$
\newcount\modularform\modularform=\equno
The requirement that $\Gu$ commutes with the RG flow means that, at a critical point $\sigma_c$,
$\beta(\gamma(\sigma_c))=\beta(\sigma_c)$, [\the\BD], (this is a generalisation to the
non-Abelian group $\Gu$ of a similar condition for the compatibility of the ${\bf Z}_2$
Kramers-Wannier type duality with RG flow, analysed in
\autoref\newcount\Damgaard\Damgaard=\refno).
This can happen if $\beta(\sigma_c)=0$, but also if $\beta(\sigma_c)=\infty$. We shall
see that the latter possibility is of interest for crossover in the QH effect.
\smallskip
There is a theorem ((4.3.4) in [\the\Rankin]) that any  function 
obeying (\the\modularform) and satisfying certain reasonable meromorphicity
properties must have a special form.
Let $$f(\sigma)=-{\vartheta_3^4\vartheta_4^4\over\vartheta_2^8}, \autoeq $$
\newcount\invariant\invariant=\equno
where $\vartheta_i$, $i=1,2,3$ are Jacobi $\vartheta$-functions 
(see the appendix  for details).
Then $f(\sigma)$ is invariant under $\Gu$ (i.e. $f(\gamma(\sigma)=f(\sigma)$)
and any function obeying (\the\modularform) must be of the form
$$\beta(\sigma)={P(f)\over f^\prime Q(f)},\autoeq $$
\newcount\Theorem\Theorem=\equno
where $P$ and $Q$ are polynomials in $f(\sigma)$.
\smallskip
In the next section equation (\the\Theorem) will be taken as a starting point and
plausible explicit analytic expressions for $\sigma(s)$ will be derived, using some further
simplifying assumptions.
\bigskip
{\bf \S 4 Renormalisation Group Flow}
\bigskip
The postulated form of the $\beta$-function, (\the\Theorem), can be restricted even
more by making some further assumptions. Let $P(f)=c\prod_{i=1}^N(f-a_i)$
be a polynomial of order $N$ and $Q(f)=\prod_{j=1}^M(f-b_j)$ be a polynomial of order $M$,  where $c$, $a_i$ and $b_j$
are constants, with no $a_i$ equal to any $b_j$ so that there are no common factors
between the two polynomials. Then $\beta$ vanishes for any $\sigma_i$ such that
$f(\sigma_i)=a_i$ and diverges for any $\sigma_j$ such that
$f(\sigma_j)=b_j$. where $f(\sigma)$ is defined in (\the\invariant). From
the theory of modular forms [\the\Rankin] it is known that only the fundamental domain 
of $\Gu$ (see figure 2 and the appendix)
need be considered as all other values of $\sigma$ in the upper-half complex
plane can be reached by acting on the fundamental domain by some element, $\gamma$, of $\Gu$.
Any critical point of $\beta(\sigma)$ in the fundamental  domain has an image 
in every copy of the fundamental domain under the action of $\Gu$.
It is a property of the theory of modular forms that $f(\sigma)$ takes all possibles complex
values at least once in the fundamental domain, and hence in all copies of the domain.
\smallskip
As explained in the previous section, the fixed points of $\Gu$ must be critical points 
of the RG flow  ($\sigma_c$ is a fixed point of $\Gu$ if there exists a
non-trivial element
$\gamma\in\Gu$ such that $\gamma(\sigma_c)=\sigma_c$). The only fixed points of $\Gu$ in the
fundamental domain are $\sigma_c=0,1,{1+i\over 2}$ and the point at infinity, $i\infty$.
These four points, and all their images under $\Gu$, must be fixed points of the RG flow.
That $\sigma_c={n+i\over 2}$, for integral $n$, is a critical point has 
substantial experimental and theoretical foundation
\autoref\newcount\YSBTS\YSBTS=\refno
\autoref\newcount\HHB\HHB=\refno. Of course $\sigma=1$ is the Hall plateau
with filling factor $\nu=1$ while $\sigma=0$ is the insulating phase, both
of which are expected to be attractive fixed points. The point $\sigma=i\infty$ would
correspond to a super-conducting phase, but this is not accessible experimentally,
although it is the point about which perturbation theory is performed in 
field theory calculations (see e.g. the article by Pruisken 
in [\the\Pruiskenb]). There is no experimental evidence for
any new critical points beyond the insulating phase $\sigma=0$,
the stable Hall plateaux (all of which are obtained
from $\sigma=1$ under the action of $\Gu$) and the critical point $\sigma_c={{1+i}\over 2}$,
and its images under $\Gu$. Experiment therefore indicates that there are no new
critical points in the fundamental domain beyond those already mentioned at $0, 1, {1+i\over 2}$
and $i\infty$. This means that the only allowed values of $a_i$ and $b_j$
in $P(f)$ and $Q(f)$ are $f(0)$, $f(1)$, $f({1+i\over 2})$ and $f(i\infty)$ --- their
images under the action of any $\gamma\in\Gu$ have the same values of $f$ since
$f$ is invariant by construction.
In fact $f(0)=f(1)=0$, $f(i\infty)=-\infty$ and $f({1+i\over 2})={1\over 4}$ (see appendix).
Obviously one does not take any $a_i$ or $b_j$ to be infinite, so one is led to
the form
$$\beta=c{f^n(f-{1\over 4})^m \over f^\prime},\autoeq$$
\newcount\betamn\betamn=\equno
where $m$ and $n$ are integers.
\smallskip
It can be proven, using standard properties of Jacobi $\vartheta$-functions (see appendix,
equation (A.9))
that
$$f^\prime=-i\pi(\theta_3^4 + \theta_4^4)f.\autoeq$$
\newcount\fprime\fprime=\equno
Thus equation (\the\betamn) can be re-expressed as
$$\beta(\sigma)={ic\over\pi}{f^{n-1}(f-{1\over 4})^m\over (\theta_3^4+\theta_4^4)}.\autoeq$$
\newcount\betanmtheta\betanmtheta=\equno

The integers $n$ and $m$ can be constrained by examining various limits.
The stability of the Hall plateau at $\sigma=1$, where $f(1)=0$, demands that $\beta=0$ when $f=0$.
The points $\sigma=1$ and $\sigma=0$ are mapped onto each other under the action of $\Gu$, so 
$\beta$ also vanishes at $\sigma=0$, where $f(0)=0$ also.
Since $\vartheta_3^4\approx -1/\sigma^2$, $\vartheta_4\rightarrow 0$ and 
$f\approx -16e^{-i\pi/\sigma}$ as $\sigma\rightarrow 0$,
(A.11), it is necessary  that $n-1\ge 0$. 
\smallskip
If the assumptions being made here have any validity
at all, the limit $\sigma\rightarrow i\infty$ should have some affinity
with the weak coupling
limit of the non-linear $\sigma$-model approach to the QH effect, advocated
by Pruisken in [\the\Pruiskenb]. Quantitative comparison of the form proposed 
here with Pruisken's asymptotic form
is hampered by the fact that the functional form of $l(\tilde b)$ is not known in 
general --- indeed the assumption that the $\beta$ function is analytic is
presumably only true, if it is true at all, in a restricted class of renormalisation schemes.
In fact the form of the $\beta$-functions suggested here implies that they cannot be
the same $\beta$-functions of Pruisken's analysis in [\the\Pruiskenb] as the latter
are definitely not holomorphic as $\sigma_{xx}\rightarrow\infty$, as noted in
[\the\LutkenBurgess].
Nevertheless it may be reasonable to extract some gross features from the weak
coupling analysis. In particular
we shall assume that for $\sigma_{xx}$ large:
\smallskip
{\global\advance\equno by 1}\newcount\asymp\asymp=\equno
\item{i)} $\beta$ is finite
\item{ii)} $\sigma_{xy}$ is a constant, i.e. $\beta_{xy}=0$\hfill(\the\asymp)
\item{iii)} $\beta_{xx}\le 0$.
\smallskip
\noindent It is to be expected that these three conditions are rather general, regardless
of whether or not holomorphicity holds.
\smallskip
Now $f\rightarrow -\infty$ and $\vartheta_3^4=\vartheta_4^4=1$ as $\sigma\rightarrow i\infty$,
so i) above requires that $n+m-1\le 0$. Since $n\ge 1$ we thus have $m\le 0$.
In perturbation theory  condition ii) comes from
the assumption that instanton effects are negligible in the weak coupling limit and
therefore $\sigma_{xy}$, which is a topological parameter in the field theory model
of Pruisken [\the\Pruiskenb], does not run in this limit.
Using the asymptotic form (A.10) for large $\sigma_{xx}$ in equation
(\the\betanmtheta) gives
$$\beta={c\over 2\pi}\left({-e^{2\pi\sigma_{xx}}\over 256}\right)^{n+m-1}
\bigl[\sin\{2\pi(n+m-1)\sigma_{xy}\} +i\cos\{2\pi(n+m-1)\sigma_{xy}\}\bigr].\autoeq$$
Condition ii) above then demands that $n=-m+1$.
Finally condition iii) requires that $c$ be real and negative.
Without loss of generality $c=-1$, by rescaling $s$ if necessary, and
one is finally led to the form
$$\beta(\sigma)=-{1\over f^\prime}{f^n\over\left(f-{1\over 4}\right)^{n-1}}=
-{i\over\pi}{1\over \left( \vartheta_3^4+\vartheta_4^4\right)}
\left({f\over f-{1\over 4}}\right)^{n-1},\autoeq$$
\newcount\betanoverm\betanoverm=\equno
with $n\ge 1$. The function (\the\betanoverm) with $n=0$ is actually the form
relevant to N=2 super-symmetric Yang-Mills theory.
\smallskip 
Equation (\the\betanoverm) can now be integrated exactly along any single trajectory. Writing it as
$${ds\over d\sigma}
=-{df\over d\sigma}{1\over f^n}\sum_{r=0}^{n-1}\pmatrix{n-1\cr r\cr}
\left(-{1\over 4}\right)^{n-1-r}f^r\autoeq$$
this integrates to
$$s(\sigma)
=-\ln\left({f\over f_0}\right)-\sum_{r=0}^{n-2}\left({1\over r-n+1}\right)\pmatrix{n-1\cr r\cr}\left(-{1\over 4}\right)^{n-1-r}
\left(f^{r-n+1}-f_0^{r-n+1}\right),\autoeq$$
where $f_0=f(\sigma(s=0))$, which can be inverted to give $\sigma(s)$.
For crossover between $\nu=0$ and $\nu=1$ we take $f_0={1\over 4}$ 
with $s({1+i\over 2})=0$.
Using the limiting form (A.11) for $f(\sigma)$ as $\sigma\rightarrow 0$,
$f\approx -16e^{-{i\pi\over\sigma}}$, there is
a quantitative difference between $n=1$ and $n>1$ in the limit of large $s$. 
In the former case, the sum
is absent and only the logarithm survives, giving $\sigma\approx i\pi/s$,
while in the latter case the term in the sum
with $r=0$ dominates and $s\approx \bigl({1\over n-1}\bigr)
\left({1\over 64}\right)^{n-1}e^{i\pi(n-1)/\sigma}$ so $\sigma\approx i\pi(n-1)/\ln s$.
The case $n=1$ therefore approaches the stable fixed points at $\sigma=0$ and $1$
for large $s$ much faster than for $n>1$. Although $s$ has not yet been determined it is
assumed to be a monotonic function of $\Delta B$ so, whatever its functional form,
$n=1$ will give faster convergence to Hall plateaux than $n>1$.
Experimentally, the former possibility is to be preferred over the latter.
\smallskip
To recapitulate, equation (\the\betanoverm) is the most general form of the $\beta$-function
compatible with the following criterion:
\bigskip
\item{i)} Stability of Hall plateaux.
\smallskip
\item{ii)} Gross features of weak coupling in the asymptotic limit of the field theoretical models.
\smallskip
\item{iii)} No new critical points of the RG flow other than the insulating phase,
Hall plateaux, critical
points in the crossover between Hall plateaux and the weak coupling  limit.
\smallskip\item{iv)} Symmetry of QH states under the action of $\Gu$ on the 
upper-half complex conductivity plane which commutes with the RG flow.
\smallskip
\item{v)} Complex analyticity of the $\beta$-functions.
\smallskip
\noindent Finally the case $n=1$ gives the fastest approach to the stable fixed
points corresponding to Hall plateaux. 
The RG flow for the case $n=1$  was considered in
[\the\LutkenLatorre] 
within the context of Yang-Mills gauge theory. It is is reproduced
in figure 1 here, which was obtained by observing that integrating (\the\betanoverm) with $n=1$
using an arbitrary integration constant gives
$$s-s_0 = -\ln\left({f\over f_0}\right),\autoeq $$
\newcount\lnform\lnform=\equno
where $f_0=f(s_0)$. Now $s$ is a real parameter, therefore $arg(f)=arg(f_0)$, and
the phase of the function $f$ in equation (\the\invariant) is constant along RG
trajectories. Figure 1 is simply a contour plot of the complex phase of $f(\sigma)$,
and it reproduces the flows predicted in [\the\Pruiskena] and [\the\Khem].
Note that $\sigma_c={1+i\over 2}$ is a repulsive fixed point.
For most of the following the case $n=1$ will be assumed, except where explicitly
stated.

\bigskip
{\bf \S 5 The Critical Point at $\sigma_c = {1+i\over 2}$}
\bigskip
The element $\gamma=\pmatrix{1&-1\cr2&-1\cr}$ of $\Gu$ leaves the point  
$\sigma_c={1+i\over 2}$ fixed and so $\sigma_c$ must be a critical point of the RG
flow, by the arguments of \S 3. $\sigma_c$ therefore corresponds to the critical
point in the crossover from $\sigma=0$ to $\sigma=1$. 
Numerical calculations, based on models for the microscopic physics,
support this conclusion [\the\HHB].
It is shown in the appendix that
$\vartheta_3^4=-\vartheta_4^4$ and $f={1\over 4}$ at $\sigma=\sigma_c={1+i\over 2}$. Since $n\ge 1$
it is immediately apparent from (\the\betanoverm) that $\beta(\sigma_c)$ diverges.
This is a critical point that is characterised not by a vanishing $\beta$-function
but by a singular one. However the situation is not as bad as it might seem
at first sight. For $\sigma$ close to the critical point, $\sigma=\sigma_c+\varepsilon$ 
with $\varepsilon$ small,
$$f\left(\sigma_c+\varepsilon\right)={1\over 4}-{\left\{\Gamma(1/4)\right\}^8\over 64\pi^4}
\varepsilon^2+o(\varepsilon^4),\autoeq$$
\newcount\fcrit\fcrit=\equno
so
% $$\vartheta_3^4 + \vartheta_3^4 = - i{\left\{\Gamma(1/4)\right\}^8\over 32\pi^4}\varepsilon
% +o(\varepsilon^2),\autoeq$$
$$f^\prime\vline_{\sigma_c}=-{\left\{\Gamma(1/4)\right\}^8\over 32\pi^4}\varepsilon
+o(\varepsilon^3),\autoeq$$
see (A.14) ($\Gamma(1/4)$ here is the usual Gamma function).
So the mildest form of singularity is
$${d\varepsilon\over ds}\approx {8\pi^4\over{\left\{\Gamma(1/4)\right\}^8}\varepsilon},\autoeq$$
\newcount\betasing\betasing=\equno
for $n=1$. This integrates to 
$\varepsilon^2\approx{16\pi^4s\over{\left\{\Gamma(1/4)\right\}^8}}$, where the zero of $s$
has been chosen so that $\sigma(s=0)=\sigma_c$. Thus $\varepsilon$ is either purely
real or purely imaginary, depending on whether $s>0$ or $s<0$. Any other value
is repulsed from $\sigma_c$, as it is an unstable fixed point. 
\smallskip 
For $\varepsilon$ real 
the choice $n=1$ thus leads to the form $\varepsilon\propto\sqrt s$,
$\Re(\beta)\propto{1\over\sqrt s}$ and $\Im(\beta)\approx 0$,
with $\sigma(s)$ itself finite at $s=0$. Only the slope of $\sigma_{xy}$ is singular
at the critical point. In fact a glance at the experimental data for crossover
between any two QH states with $\sigma_{xy}=\nu_1$ and $\sigma_{xy}=\nu_2$
makes it clear that one {\it expects} the slope of $\sigma_{xy}(B)$ to be
large (and indeed infinite as the temperature $T\rightarrow 0$, where one expects
a step function). Thus this singularity is not necessarily a disaster.
For finite temperatures, the infinite slope can be avoided
by a suitable choice of $s(\Delta B)$ or, equivalently, $s(\Delta\nu)$.
Recall that $s(\Delta B)$ is taken to be a monotonic function of the
external magnetic field $B$ and one is free to choose the zero of $s$ so that
$s(\Delta B=0)=0$, or $s(\Delta\nu=0)=0$. We assume that $s(\Delta\nu)$ is
a real analytic function of $\Delta\nu$ at $\Delta\nu=0$, otherwise
the definition of the $\beta$-function in (\the\nabeta) hardly makes 
sense.\footnote*{I thank Denjoe O'Connor for discussions on this point.}
Experimentally [\the\Shaharetal] $\sigma_{xx}$ and $\sigma_{xy}$ are perfectly well behaved
functions of $\Delta\nu$ near $\Delta\nu=0$, they look analytic with
${d\sigma_{xy}\over d(\Delta\nu)}>0$ and finite, and ${d\sigma_{xx}\over d(\Delta\nu)}=0$.
If we therefore assume a Taylor expansion
$$\sigma_{xy}(\Delta\nu)={1\over 2}+C\Delta\nu+\cdots,\autoeq$$
with $C$ a positive constant, we see that $\Delta\nu\propto\varepsilon\propto\sqrt s$,
i.e. $s\propto(\Delta\nu)^2$. This gives rise to a singularity in 
${d\sigma_{xy}\over d s}$ at $s=0$, even though 
${d\sigma_{xy}\over d(\Delta\nu)}$ is perfectly finite for finite temperature.
This is the source of the pole in the $\beta$-function at $\sigma_c={1+i\over 2}$.
\smallskip 
For general $n\ge 1$ a similar analysis shows
that   ${d\epsilon\over ds}\propto{1\over\epsilon^{2n-1}}$ 
near the critical point, giving $\epsilon\propto s^{1/(2n)}$
and  $\Re(\beta)\propto s^{-({2n-1\over2n})}$. Analyticity of $s(\Delta\nu)$
at the critical point then requires $s\approx (\Delta\nu)^{2n}$.
From now on only $n=1$ will be considered as it appears likely to be the
most relevant in order to achieve the most rapid approach to the fixed
points on the real axis.
\smallskip
\bigskip
{\bf \S 6 Analytic form of the Crossover}
\bigskip
In this section we consider the explicit analytic form of the function $\sigma(s)$,
for $n=1$, in
crossing over from $\sigma=0$ to $\sigma=1$ along the semi-circular arch of unit
diameter passing through $\sigma={1+i\over 2}$. All QH transitions, between
$\sigma=\nu_1=p_1/q_1$ and $\sigma= \nu_2=p_2/q_2$ with $p_2q_1-p_1q_2=1$, can
be obtained by mapping this semi-circle onto the semi-circle bridging $\nu_1$ and $\nu_2$,
with diameter ${1\over q_1q_2}$ passing through,
$\gamma(\sigma_c)={p_1q_1+p_2q_2+i\over q_1^2+q_2^2}$. We have, from (\the\lnform),
$$f(\sigma)=f_0\e^{-(s-s_0)}.\autoeq$$
Choosing the zero of $s$, $s_0$, so that $s=0$ corresponds to the point $\gamma(\sigma_c)$,
where $f(\gamma(\sigma_c))=f_0=1/4$ results in
$$-{\vartheta_3^4\vartheta_4^4\over\vartheta_2^8}=f(\sigma)={1\over 4}\e^{-s}.
\autoeq$$
\newcount\expform\expform=\equno
Thus the semi-circular arches giving crossover between any two QH states are
characterised by $arg(f)=0$. The explicit functional form of $\sigma(s)$ can now
be determined in terms of complete elliptic functions of the second kind,
$$K(k)=\int_0^{\pi/2}{d\varphi\over\sqrt{1-k^2\sin^2\varphi}},\autoeq $$
using standard formulae
% For some reason \refno has got corrupted. So I set it
% back to its last correct value manually!
%{\global\refno=\HHB}\kern -1pt
\autoref\newcount\WW\WW=\refno (see appendix),
$$\vartheta_2=\sqrt{2kK(k)\over\pi},\qquad
\vartheta_3=\sqrt{2K(k)\over\pi},\qquad
\vartheta_4=\sqrt{2k^\prime K(k)\over\pi},\autoeq$$
\newcount\thetaK\thetaK=\equno
where the modulus, $k$, is related to $\sigma$ by $\e^{i\pi\sigma}=\e^{-\pi K^\prime/K}$,
with $K^\prime(k)=K(k^\prime)$ and \hbox{$(k^\prime)^2=1-k^2$} the complementary modulus.
Thus
$$\sigma=2\bar m+{iK^\prime\over K},\autoeq $$
\newcount\sigmas\sigmas=\equno
for integral $\bar m$ (for convenience the relevant formulae are reproduced in the
appendix).
The integer $\bar m$ can be set to zero, since the other possibilities follow from 
repeated application of the
symmetry $\sigma\rightarrow\sigma+1$ of $\Gu$.
Using the relations (\the\thetaK) in (\the\expform) gives $s$ in terms of $k$,
$${1\over 4}\e^{-s}=-{\vartheta_3^4\vartheta_4^4\over\vartheta_2^8}=-{(k^\prime)^2\over k^4}
\qquad\Rightarrow\qquad k^2=2\e^s\left(1\pm\sqrt{1-e^{-s}}\right).\autoeq $$
\newcount\ks\ks=\equno
\smallskip
For $0\le s<\infty$ the plus sign gives $2\le k^2<\infty$ and the negative sign gives
$1<k^2\le 2$ --- the two branches combined give the range $1<k^2<\infty$,
and so $-\infty<(k^\prime)^2<0$
(it has already been shown  --- see the discussion after equation (\the\betasing) ---
that $s<0$ corresponds to passing through $\sigma_c$ in the purely imaginary
direction, which is not relevant to crossover).
 Since
$K(k)$ is complex
in this range it is convenient to manipulate equation (\the\sigmas), using some identities
between elliptic integrals given in the appendix, in order to expose the real and 
imaginary parts of (\the\sigmas) explicitly. First note that $w^2:=1/k^2$ has
the range $0<w^2<1$, so equation (A.19) of the appendix
allows one to write
$$K(k)=w\left\{K(w)+iK^\prime(w)\right\},\autoeq $$
\newcount\Kone\Kone=\equno
with $K(w)$ and $K^\prime(w)$ real. Next define a real variable $u$ via $k^\prime=iu$,
with $0<u<\infty$, $w^2={1\over 1+u^2}$ and $(w^\prime)^2=1-w^2={u^2\over 1+u^2}$.
Then equation (A.20) of the appendix, with $k$ replaced by $w^\prime$,
gives
$$K^\prime(k)=K(iu)=wK^\prime(w).\autoeq $$
\newcount\Ktwo\Ktwo=\equno
Using equations (\the\Kone) and (\the\Ktwo) in (\the\sigmas), 
with $\bar m=0$, results in
$$\sigma(s)={K^\prime(w)\{K^\prime(w)+iK(w)\}\over [\{K(w)\}^2 + 
\{K^\prime(w)\}^2]},\autoeq $$
\newcount\sigmaw\sigmaw=\equno
which gives an explicit analytic form for $\sigma(s)$ in which the real and
imaginary parts are manifest,  since 
$$w={1\over k}=\sqrt{1\mp\sqrt{1-\e^{-s}}\over 2}\autoeq$$ from (\the\ks),
lies in the range $0<w<1$.
\smallskip
In order to show that the above equation  does indeed describe crossover between $\sigma=0$
and $\sigma=1$ we shall analyse various limits of (\the\sigmaw).
For $s=0$ both branches give $w^2={1\over 2}$ and so $\sigma(0)={1+i\over 2}$,
since $K(1/\sqrt 2)=K^\prime(1/\sqrt 2)$, thus $s=0$ is indeed the critical point.
As $w\rightarrow 0$ ($s\rightarrow\infty$ on the upper branch), the asymptotic forms
of the complete elliptic integrals (equations (A.17) and (A.18) in the appendix)
can be used to write
$$K(w)={\pi\over 2} + o(w^2) = {\pi\over 2}+o(\e^{-s}), \autoeq $$
$$ K^\prime(w)=\ln\left({4\over w}\right) + o(w^2\ln w) = {s\over 2}+\ln 8 +o(s\e^{-s}),\autoeq $$
so
$$\sigma={(s+6\ln 2)(s+6\ln 2+i\pi)\over\{\pi^2+(s+6\ln 2)^2\}} + o(s\e^{-s})
\quad\rightarrow\quad 1.\autoeq $$
As $w\rightarrow 1$ ($s\rightarrow\infty$ on the lower branch), the r\circum{o}les of
$w$ and $w^\prime$ are interchanged and
$$ \sigma={\pi\{\pi+i(s+6\ln 2)\}\over\{\pi^2+(s+6\ln 2)^2\}} + o(s\e^{-s})
\quad\rightarrow\quad 0.\autoeq$$
\newcount\PH\PH=\equno
Thus the two branches combined, with $0<w<1$, do indeed interpolate between
$\sigma=0$ and $\sigma=1$ with $w=1/\sqrt 2$ giving the critical point at
$\sigma_c={1+i\over 2}$.
\smallskip
To compare this analytic expression for $\sigma(s)$ with the
available experimental data, we must still address the question of how the
variable $s$ is depends on the external magnetic field, or equivalently
on $v=\bigl({\bar n e\over T^\mu}\bigr)\Delta\nu$. 
One piece of information that has not yet been brought into play,
and can be used to constrain the form of
$s$, is the particle-hole transformation rule of [\the\KLZ].
This implies a symmetry under $\nu\rightarrow 1-\nu$, or 
$\Delta\nu\rightarrow -\Delta\nu$, which suggests that $s$ should
be an {\it even} function of $\Delta\nu$.\footnote*{We shall see later that
this is strictly true only for integer transitions.}
This is compatible with the mathematical analysis 
in the previous section, where it was shown that the apparent regularity 
in the experimental data of $\sigma_{xy}$ at the critical point
implies that $s\propto v^2$,
at least for $\Delta\nu$ small. 
As a first try, therefore, we shall take
$s=(A\Delta\nu/T^\mu)^2$, for some real positive constant $A$. 
It may be significant that the experimental value of the critical exponent 
in (\the\Bscale), $\nu_\xi\approx 2.02$, is so close to two. If it were precisely
two, then $s$ would be proportional to the inverse correlation
length, $s\propto 1/\xi$. It would also be quite natural if $\mu$
were exactly one half, giving $s\propto 1/T$. The experimental data, however, seem
to indicate a somewhat lower value for $\mu$ (and correspondingly a higher value
for $\nu_\xi$).
\smallskip
Using $s=\left({A\Delta\nu\over T^\mu}\right)^2$ and
the four values for the temperature quoted for figure 2b in
[\the\Shaharetal], $T=42$, $84$,
$106$ and $145$ mK, with 
the best fit experimental value of $\mu=0.45\pm0.05$, the functions
$$\sigma_{xy}(s)={\{K^\prime(w)\}^2 \over [\{K^\prime(w)\}^2 + \{K(w)\}^2]},$$
$$\sigma_{xx}(s)={K^\prime(w)K(w)\over [\{K^\prime(w)\}^2 + \{K(w)\}^2]},\autoeq $$
\newcount\sigmawtemplate\sigmawtemplate=\equno
can be plotted, with
$$w=\sqrt{1-\hbox{sign}(\Delta\nu)
\sqrt{1-\e^{-\left({A\Delta\nu\over T^\mu}\right)^2}}\over 2},
\autoeq $$
\newcount\wBone\wBone=\equno
for the transition $\nu:0\rightarrow 1$, 
and the result for $\rho_{xx}$ and $\rho_{xy}$ are shown in figure 3,
where the choices $A=60$, $\mu=0.50$ have been made, since these seem
to give a good visual fit to the experimental data in figure 2b
of [\the\Shaharetal]. A better fit to the data is obtained
if $\rho_{xx}$ is rescaled by a constant, so as to raise the
critical value of $\rho_{xx}$ above the predicted value of unity ---
the experimental curve in [\the\Shaharetal] gives a value greater than
one whereas $\Gamma_0(2)$ symmetry predicts exactly one. If
the assumptions made here are correct, this is presumably due 
to the experimental
difficulties involved in determining $\rho_{xx}$ (see for
example the comments in Cage's article in [\the\Pruiskenb]).
% For some reason \refno has got corrupted. So I set it
% back to its last correct value manually!
%{\global\refno=\WW}\kern -1pt
%\autoref\newcount\PrangeGirvin\PrangeGirvin=\refno).
Note that a prediction here is that,
for this transition, $\rho_{xy}=1$ well into the insulating phase ---
a fact that is well borne out by experiment.
Similarly, with the four values $T=42$, $70$,
$101$ and $137$ mK used in figure 1 of [\the\Shaharetal] 
the functions
$$\sigma_{xy}(s)=1+{\{K^\prime(w)\}^2 \over [\{K^\prime(w)\}^2 + \{K(w)\}^2]},$$
$$\sigma_{xx}(s)={K^\prime(w)K(w)\}\over [\{K^\prime(w)\}^2 + \{K(w)\}^2]},\autoeq $$
\newcount\sigmaxxyw\sigmaxxyw=\equno
for the transition $\nu:1\rightarrow 2$, 
are plotted in figure 4, together with the corresponding resistivities.
Figure 4 was produced with the choice $A=40$, $\mu=0.50$, again
with a view to a good visual fit to the experimental data
in figure 1 of [\the\Shaharetal], and again the fit for
$\rho_{xy}$ is noticeably better than that for $\rho_{xx}$. 
\smallskip
It is stressed that there are only two
parameters which can be varied in the analytic expressions to 
produce figures 3 and 4, the scaling exponent $\mu$ 
and the constant $A$ --- 
$\mu$ is assumed to be universal and is taken from experiments, 
which then leaves only one parameter, $A$,
which can be varied in order to fit the experimental data. 
In general $A$ is not expected to be universal --- it would
depend on various parameters such as the electron and
impurity density. In particular it depends on the critical magnetic
field and so is different for each transition.

It should also be borne in mind that the form (\the\wBone) for $w(\Delta\nu)$
could have higher corrections in $\Delta\nu$, 
it is only for $\Delta\nu$ near zero that we can trust
$s\propto\left({\Delta\nu\over T^\mu}\right)^2$.
However, assuming that $s$ is an even function of $\Delta\nu$,
as the particle-hole transformation mentioned earlier suggests,
one expects the corrections to be of order $(\Delta\nu)^4$.
In the experimental data from reference [\the\Shaharetal], 
the range of $\Delta\nu$
required for crossover is $<0.1$ so corrections to the 
leading $(\Delta\nu)^2$ term can be expected
to be less than $1\%$, unless something conspires
to produce large co-efficients.
\smallskip
The particle-hole transformation extends to complex $\sigma$
as $\sigma\rightarrow1-\bar\sigma$, as is evident from
figure 2, (this is not an element of $\Gamma_0(2)$, but rather
an outer auto-morphism of that group).
Thus it is to be expected that the longitudinal conductivity,
$\Im\{\sigma(\Delta\nu)\}$,
is an even function of $\Delta\nu$, at
least for integral transitions.
This property is manifest in 
equations (\the\sigmawtemplate) and (\the\sigmaxxyw)
since, as is easily proven from (\the\wBone),
$w(-\Delta\nu)=w^\prime(\Delta\nu)$ where $(w^\prime)^2=1-w^2$. 
%Therefore any corrections 
%would be of order $(\Delta\nu)^4$ or higher, 
%but equation (\the\wBone) appears to fit the experimental data
%quite well as it stands.
\smallskip
This behaviour of $w(\Delta\nu)$ under $\Delta\nu\rightarrow-\Delta\nu$
has an interesting
consequence for the resistivity, $\rho=-1/\sigma$, derived from
equation (\the\sigmawtemplate).
The longitudinal resistivity for the transition $\nu:0\rightarrow 1$ is
$$\rho_{xx}={K(w)\over K^\prime(w)}.\autoeq $$
\newcount\dualrho\dualrho=\equno
Thus we have 
$$\rho_{xx}(-\Delta\nu)={1\over \rho_{xx}(\Delta\nu)},\autoeq $$
\newcount\rhoduality\rhoduality=\equno
for $\nu:0\rightarrow 1$,which is well supported experimentally
[\the\STSCSS]
% For some reason \refno has got corrupted. So I set it
% back to its last correct value manually!
{\global\refno=\WW}\kern -1pt
\autoref\newcount\SHLTSR\SHLTSR=\refno
\autoref\newcount\HSSTXM\HSSTXM=\refno.
In particular in reference [\the\SHLTSR]
the authors fit the form 
$\rho_{xx}(\Delta\nu)=\e^{-{\Delta\nu\over\tilde\nu(T)}}$
to the data, where $\tilde\nu(T)$ is a function of $T$.
They find that $\tilde\nu(T)=T^\mu$ is incompatible with the data
for any $\mu$ and suggest instead that a linear form,
$\tilde\nu(T)=\tilde\alpha T+\tilde\beta$
with $\tilde\alpha$ and $\tilde\beta$
non-zero constants, gives a better fit and this is interpreted
as a violation of the scaling hypothesis. It would be
very interesting to check whether or not the alternative
form (\the\dualrho) is compatible with the experimental
data because, if it is, then the experiments in
[\the\STSCSS] [\the\SHLTSR] [\the\HSSTXM] 
would provide confirmation of scaling
and not a violation of it.
\smallskip
Taking (\the\sigmaw) as a template, the assumption of $\Gu$ symmetry (which
implies universality) allows any transition to be modeled. The crossover
$\nu:{p_1\over q_1}\rightarrow{p_2\over q_2}$ is obtained from the template transition
$\nu:0\rightarrow 1$ by the action of $\gamma=\pmatrix{p_2-p_1&p_1\cr q_2-q_1&q_1\cr}$
as (see appendix)
$$\sigma\rightarrow{(p_2-p_1)\sigma+p_1 \over (q_2-q_1)\sigma +q_1}.\autoeq$$
Using (\the\sigmaw) this gives 
$$\sigma(\Delta\nu)={p_2q_2\{K^\prime(w)\}^2+p_1q_1\{K(w)\}^2
+ iK^\prime(w)K(w) \over [\{q_1^2K(w)\}^2 + q_2^2\{K^\prime(w)\}^2]}.\autoeq $$
\newcount\crosspq\crosspq=\equno
The form of $w$ as a function of $\Delta\nu$ changes as well, because 
$\Delta\nu$ for the transition $\nu:{p_1\over q_1}\rightarrow{p_2\over q_2}$
is not the same as $\Delta\nu$ for the template transition $\nu:0\rightarrow 1$,
in general. Denoting the filling factor for the template transition by $\nu_{01}$
(which is the argument of the function $s(\Delta\nu_{01}$)),
we expect $\nu_{01}$ to be obtained from $\nu$ using 
$\gamma^{-1}=\pmatrix{q_1&-p_1\cr -(q_2-q_1)&p_2-p_1\cr}$ so
$$\nu_{01}={q_1\nu-p_1 \over (q_1-q_2)\nu +(p_2-p_1)},\autoeq $$
and thus
$$\Delta\nu_{01}={\Delta\nu\over 
\alpha\{(q_1-q_2)\Delta\nu+\alpha\}}:={\Delta\nu\over\zeta(\Delta\nu)}\autoeq $$
\newcount\nupq\nupq=\equno
where $\alpha:=(p_2-p_1)-(q_2-q_1)\nu_c$ and 
$\zeta(\Delta\nu):=\alpha\{(q_1-q_2)\Delta\nu+\alpha\}$.
$\Delta\nu$ and $\Delta\nu_{01}$ are equal for integer transitions
of the form $p\rightarrow p+1$ (where $p_2=p+1$, $p_1=p$ and $q_1=q_2=1$) but not otherwise.
Note that $\alpha\rightarrow -\alpha$, $q_1\rightarrow q_1$ and 
$q_2\rightarrow q_2$ under $\nu\rightarrow 1-\nu$, hence $\Delta\nu_{01}\rightarrow-\Delta\nu_{01}$
and the discussion of the particle-hole transformation rule, in the paragraph
after equation (\the\PH), should now be modified to say that
$s$ could have corrections of order $(\Delta\nu_{01})^4$.
\smallskip
Thus for a general crossover the argument $w$ appearing in (\the\crosspq )
is given by 
$$w=\sqrt{1-\hbox{sign}(\Delta\nu)
\sqrt{1-\e^{-
\left({A\Delta\nu\over\zeta(\Delta\nu)T^\mu}\right)^2}}\over 2}.
\autoeq $$
\newcount\wbtwo\wbtwo=\equno
Equation (\the\crosspq), with the definition (\the\wbtwo),
is the main result of this paper. 
In principle there could be corrections of order $(\Delta\nu)^4$
to the exponent in equation (\the\wbtwo), but in practice it appears to give
good agreement with experiment as it stands, since the range of
$\Delta\nu$ required for crossover is small.
The special case  (\the\sigmaw)
is reproduced by setting $p_1=0$ and $p_2=q_1=q_2=1$.
\smallskip
The longitudinal resistivity actually takes the form given in (\the\dualrho)
for all of the
transitions $\nu:0\rightarrow 1/q$, as can be shown using (\the\crosspq)
with $p_1=0$, $p_2=q_1=1$ and $q_2=q$.
Thus \lq\lq rho-duality'',(\the\rhoduality), should be valid for these transitions, but only these.
Equation (\the\rhoduality), with $\Delta\nu$ replaced with $\Delta\nu_{01}$ from (\the\nupq),
was tested experimentally in [\the\STSCSS] for
the case $\nu:0\rightarrow 1/3$ 
(for which $p_1=0$, $p_2=q_1=1$ and $q_2=3$) 
and appears to give support for (\the\nupq), even for values of $\Delta\nu$
as large as $0.1$, while putting $\zeta(\Delta\nu)$ equal to a constant
does not give such good a fit to the data, the discrepancy
being a $10\%$ correction. It therefore seems that (\the\wbtwo)
is valid even for $\Delta\nu$ as large as $0.1$, at least for the temperature
range explored in [\the\STSCSS], with the corrections being of order
$(\Delta\nu)^4$ giving a $1\%$ error which is within the limits of
experimental accuracy.
\bigskip
{\S \bf 7 Conclusions}
\bigskip
In this paper an explicit form of the crossover between two QH plateaux has
been derived, the final result being given in (\the\crosspq) and (\the\wbtwo).
The resulting crossovers 
are plotted in figures 3 and 4 for the cases $p_1=0$, $p_2=q_1=q_2=1$ and
$p_2=2$, $p_1=q_1=q_2=1$, corresponding
to $\nu:0\rightarrow 1$ and $\nu:1\rightarrow 2$ respectively. 
The results agrees remarkably well with the experimental
data in [\the\Shaharetal], at least qualitatively. It would be of interest to check the results
quantitatively, using experimental numbers rather than just the graphical data
which are available in the literature. 
\smallskip The assumptions are:
\bigskip
\item{i)} The law of corresponding states of [\the\KLZ] 
should be extended into the upper-half complex conductivity plane,
$\sigma=\sigma_{xy}+i\sigma_{xx}$ (in units of $e^2/h$), as
required by rotational invariance in the bulk,. This is
encoded mathematically into a group action on $\sigma$ --- specifically
the group is $\Gu$.
\smallskip
\item{ii)} The action of $\Gu$ commutes with RG flow. This implies
universality of QH transitions.
\smallskip
\item{iii)} There are no extra critical points, other than those already
indicated by the experimental data and weak coupling expansions, i.e.
$\sigma=0,1,{1+i\over 2},i\infty$ and their images under the action of $\Gu$,
and that the RG flow be compatible with the experimental data and the robust
features of the weak
coupling expansion enumerated in (\the\asymp).
\smallskip
\item{iv)} The $\beta$-functions for the RG flow that describes  crossover
between two QH plateaux, or between a QH state and the insulating phase,
are complex analytic functions when described in terms of a real
variable $s$ that is a monotonic real analytic function of the external
magnetic field. 
\bigskip
Assumptions i) and ii) above have already received a great deal of attention,
[\the\LutkenRossb] [\the\Lutken] [\the\BD] and [\the\KLZ]. Despite many successes
it is still not yet clear whether universality in QH transitions is a good hypothesis,
but time will
tell and for the moment it seems promising to investigate the
consequences of these assumptions.
Assumption iii) does not seem to require comment, given i) and ii). 
The new ingredient here is iv)
which necessitates analysis and justification. The only justification given
here is that this assumption produces an expression which appears to be in
remarkable agreement with experimental data on QH crossover. Given this apparent
success, it is important to achieve an understanding of how this form could emerge from
the underlying physics. There is no obvious reason from the field theory
models considered so far, e.g. Pruisken's non-linear $\sigma$-model with a topological
term [\the\Pruiskenb] or Chern-Simons effective actions 
\autoref\newcount\Baletal\Baletal=\refno, why an analytic $\beta$-function should 
give good results.
It should probably be borne in mind when trying to analyse the hypothesis
of complex analyticity using field theoretic techniques that the notion
of complex analyticity would be expected to depend on the renormalisation scheme
chosen for any calculations --- it could not be expected to be true in an
arbitrary scheme. It may be significant for analyticity
that if the exponent $\nu_\xi$ in
$(\the\Bscale)$ were exactly two, then the RG parameter $s$ would be inversely
proportional to the correlation length.
\smallskip
Finding a microscopic justification 
promises to be an involved project, but in view of the tantalising
similarity between the analytic results presented
here and the experimental data it may be one well worth pursuing.
\bigskip
It is a pleasure to thank Jan Pawlowski for helpful discussions during the
early stages of this work and Jason Twamley for help in producing
figure 1.
\smallskip
This research was sponsored in part by an Alexander von Humboldt foundation fellowship
and in part by financial assistance from
Baker Consultants Ltd., Ireland, networking specialists
(http://www.baker.ie).
\bigskip
{\bf Appendix}
\bigskip
This appendix contains a summary  of the properties of the modular group, its sub-group
$\Gu$, the Jacobi $\vartheta$-functions and complete integrals of the second kind,
which are needed in the text.
\smallskip
The modular group, $\Gamma(1)$, can be represented by the set of all $2\times 2$
matrices with integer entries and determinant one. Thus $g\in\Gamma(1)$ can be
written as
$$g=\pmatrix{a&b\cr c&d\cr} \qquad\hbox{with}\qquad ad-bc=1.\autoeqA{A}$$
$\Gamma(1)$ is therefore isomorphic to $Sl(2,{\bf Z})$,  which in turn is isomorphic to
$Sp(2,{\bf Z})$ --- the group of symplectic $2\times 2$ matrices with integer entries.
Elements of $\Gamma(1)$ have a natural action on the upper-half complex plane,
parameterised here by $\sigma\in{\bf C}$ with $\Im(\sigma)>0$, given by
$$g(\sigma)={a\sigma+b\over c\sigma+d}.\autoeqA{A}$$
\newcount\mobiusA\mobiusA=\equnoA
It is easy to check, using $ad-bc=1$, that $\Im\bigl(g(\sigma)\bigr)>0$ if $\Im(\sigma)>0$.
The group $\Gamma(1)$ is an infinite discrete group  and it is generated  by two elements,
$U:\sigma\rightarrow \sigma+1$ with  $g_U=\pmatrix{1&1\cr 0&1\cr}$ and 
$V:\sigma\rightarrow -{1\over\sigma}$ with $g_V=\pmatrix{0&1\cr -1&1\cr}$, any element
can be represented by some combination of products of such matrices (though not
necessarily uniquely).
The group $\Gamma(1)$ is a discrete version
of the group of $2\times 2$ matrices with real entries and determinant one --- the
special linear group, $Sl(2,{\bf R}$), which has a similar action on the upper-half complex
plane, (A.\the\mobiusA). With this action $Sl(2,{\bf R})$ maps semi-circles
centred on the real line onto other semi-circles with the same
property and obviously the discrete sub-group $\Gamma(1)$ must share this feature.
\smallskip
Clearly $\Gamma(1)$ maps rational numbers on the real line $\sigma={p\over q}$
to other rational numbers $\sigma={ap+dq\over cp+dq}$. A general element of
$\Gamma(1)$ does not necessarily preserve the parity (even or odd) of the denominator
under such action,
$cp+dq$ can be either even or odd regardless of the parity of $q$, and
so $\Gamma(1)$ itself can hardly be a symmetry group between QH states, 
$\sigma=\nu={p\over q}$, for
which $q$ must be odd. However the sub-group of $\Gamma(1)$ whose bottom left entry
is even does preserve the parity of the denominator, and the possible significance of this 
property for the QH effect was first noticed by L\"utken and Ross in [\the\LutkenRossb].
This sub-group is often denoted by $\Gu$ in the mathematical literature (though
reference [\the\Rankin] denotes it by $\Gamma_U(2)$) and it can be represented by
matrices of the form
$$\gamma=\pmatrix{a&b\cr 2c&d\cr} \qquad\hbox{with}\qquad ad-2bc=1,\autoeqA{A}$$
with $a$, $b$, $c$ and $d$ integers.
$\Gu$ is generated by two elements
$U:\sigma\rightarrow \sigma+1$ with  $g_U=\pmatrix{1&1\cr 0&1\cr}$ and 
$X:\sigma\rightarrow{\sigma\over 2\sigma+1}$ with $g_X=\pmatrix{1&0\cr 2&1\cr}$.
These give a generalisation, [\the\BD], to the whole upper-half complex plane
of the Landau level addition transformation 
and the flux attachment rule respectively of Kivelson, Lee and Zhang, [\the\KLZ],
which were originally defined for rational
filling factors only, corresponding to QH states.
\smallskip
This action of $\Gu$ on the upper-half $\sigma$-plane results in an inhomogeneous
tiling of the plane --- the whole plane can be generated by the action of $\Gu$
on a \lq\lq fundamental domain'', which can be taken to be a connected domain,
[\the\Rankin]. The definition  of the fundamental domain is not unique, 
but a convenient choice is a vertical strip of unit radius extending infinitely
far in the imaginary direction, but with its lower edge bounded by a semi-circular
arch of unit diameter connecting the points $\sigma=0$ and $\sigma=1$ (see figure 2).
The properties of any function are uniquely determined by knowledge of the
function on the fundamental domain alone.
\smallskip
Since $\Gu$ is a sub-group of the full modular group it  maps semi-circles 
centred on the real line onto other semi-circles with the same
property. In particular the above mentioned semi-circular arch of radius ${1\over 2}$ spanning
$\sigma=0$ and $\sigma=1$ is mapped by $\gamma=\pmatrix{a&b\cr 2c&d}$ into an arch
of radius ${1\over 2d(2c+d)}$, spanning $\sigma={b\over d}$ and $\sigma={a+b\over 2c+d}$.
If ${b\over d}={p_1\over q_1}$ and ${a+b\over 2c+d}={p_2\over q_2}$, with $q_1$ and 
$q_2$ odd, are two QH states between which a transition is allowed we have
$a=p_1-p_2$, $b=p_1$, $2c=q_2-q_1$ and $d=q_1$ so $\gamma=\pmatrix{p_2-p_1&p_1\cr q_2-q_1&q_1\cr}$.
Now the condition $\hbox{det}\gamma=1\iff p_2q_1-p_1q_2=1$ which gives a
selection rule for QH transitions, [\the\BD].
In this way all allowed QH transitions can be generated from a \lq\lq fundamental''
transition describing crossover between the the insulating state, with $\sigma=0$, and the 
QH state with filling factor unity, with $\sigma=1$, $\nu:0\rightarrow 1$,
by the action of some element of $\Gu$. This immediately implies the hypothesis of universality
of QH transitions [\the\Pruiskena]. 
\smallskip
The fixed points of $\Gu$ play an important r\circum{o}le in crossover phenomena.
The point $\sigma_c={1+i\over 2}$ is left invariant by the element 
$\gamma_c=\pmatrix{1&-1\cr 2&-1\cr}$, and there exist elements of $\Gu$ which leave
any image, $\gamma(\sigma_c)$, of $\sigma_c$ fixed, namely 
$\gamma\gamma_c\gamma^{-1}\{\gamma(\sigma_c)\}=\gamma(\sigma_c)$.
Under the assumption that the action of $\Gu$ on the $\sigma$-plane commutes with 
the RG flow, $\sigma_c$ and all its images must be fixed points of the RG flow,
[\the\LutkenRossb] [\the\Lutken] [\the\BD].
\smallskip
The Jacobi functions $\vartheta$-functions used in the text are defined by
$$\eqalign{\vartheta_2&=2\sum_{n=0}^\infty q^{(n+{1\over 2})^2}=
2q^{1\over 4}\prod_{n=1}^\infty\bigl(1-q^{2n}\bigr)\bigl(1+q^{2n}\bigr)^2,\cr
\vartheta_3&=\sum_{n=-\infty}^\infty q^{n^2}=
\prod_{n=1}^\infty\bigl(1-q^{2n}\bigr)\bigl(1+q^{2n-1}\bigr)^2,\cr
\vartheta_4&=\sum_{n=-\infty}^\infty (-1)^nq^{n^2}=
\prod_{n=1}^\infty\bigl(1-q^{2n}\bigr)\bigl(1-q^{2n-1}\bigr)^2, \cr}\autoeqA{A} $$
\newcount\thetadef\thetadef=\equnoA
where $q:=\e^{i\pi\sigma}$ (the conventions are those of [\the\WW], except that $\tau$
there is replaced by $\sigma$ here).
\smallskip
The $\vartheta$-functions satisfy the relation
$$\vartheta_3^4=\vartheta_2^2+\vartheta_4^4 \autoeqA{A} $$
\newcount\thetarelation\thetarelation=\equnoA
and have the following transformations under 
$U:\sigma\rightarrow \sigma+1$ and $V:\sigma\rightarrow -{1\over\sigma}$
$$\eqalign{
\vartheta_2(\sigma+1)=\e^{i\pi\over 4}\vartheta_2(\sigma) ,
\qquad & \vartheta_2\left(-{1\over\sigma}\right)=\sqrt{-i\sigma}\;\vartheta_4(\sigma),\cr
\vartheta_3(\sigma+1)=\vartheta_4(\sigma),
\;\quad\qquad & \vartheta_3\left(-{1\over\sigma}\right)=\sqrt{-i\sigma}\;\vartheta_3(\sigma),\cr
\vartheta_4(\sigma+1)=\vartheta_3(\sigma), 
\;\quad\qquad & \vartheta_4\left(-{1\over\sigma}\right)=\sqrt{-i\sigma}\;\vartheta_2(\sigma),\cr}
\autoeqA{A}$$
\newcount\thetaUX\thetaUX=\equnoA
(a demonstration of these transformation properties can be found in  [\the\WW]).
\smallskip
It is not difficult to verify, using the above relations, that the function 
$$f(\sigma)=-{\vartheta_3^4\vartheta_4^4\over\vartheta_2^8}=
-{1\over 256q^2}\prod_{n=1}^\infty{\bigl(1-q^{4n-2}\bigr)^8\over \bigl(1+q^{2n}\bigr)^{16}}
\autoeqA{A}$$
\newcount\invariantA\invariantA=\equnoA
is invariant under the above transformations $U$ and $X$, and hence invariant under all of 
$\Gu$. 
\smallskip
The Jacobi $\vartheta$-functions satisfy the following differential equations
(see [\the\Rankin], p.231, equation (7.2.17)),
$${\vartheta_3^\prime\over\vartheta_3}-{\vartheta_4^\prime\over\vartheta_4}
={i\pi\over 4}\vartheta_2^4,\qquad
{\vartheta_2^\prime\over\vartheta_2}-{\vartheta_3^\prime\over\vartheta_3}
={i\pi\over 4}\vartheta_4^4,\qquad
{\vartheta_2^\prime\over\vartheta_2}-{\vartheta_4^\prime\over\vartheta_4}
={i\pi\over 4}\vartheta_3^4,\autoeqA{A} $$
\newcount\thetadiff\thetadiff=\equnoA
where $^\prime ={d\over d\sigma}$. Using these to differentiate (A.\the\invariantA) one finds
$${df\over d\sigma}=-i\pi\bigl(\vartheta_3^4+\vartheta_4^4\bigr)f.\autoeqA{A} $$
\newcount\fprime\fprime=\equnoA
\smallskip
In the text the following asymptotic forms of the $\vartheta$-functions
are required,which are easily verified from the definitions (A.\the\thetadef):
\smallskip
$$\hbox{i)}\qquad \sigma\rightarrow i\infty:\quad 
\vartheta_2\approx 2\;\e^{i\pi\sigma\over 4}\rightarrow 0,
\quad \vartheta_3\rightarrow 1,\quad 
\vartheta_4\rightarrow 1,
\quad f\approx-{\e^{-2\pi i\sigma}\over 256}\rightarrow -\infty.\autoeqA{A} $$
From these one can deduce, using (A.\the\thetaUX),
$$\hbox{ii)}\qquad \sigma\rightarrow 0:\quad \vartheta_2\approx\sqrt{i\over\sigma},
\quad \vartheta_3\approx\sqrt{i\over\sigma},\quad 
\vartheta_4\approx 2\sqrt{i\over\sigma}\;\e^{-{i\pi\over 4\sigma}}\rightarrow 0,
\quad f\approx -16\;\e^{-{i\pi\over\sigma}}\rightarrow 0.\autoeqA{A} $$
\smallskip
The behaviour of $f$ near $\sigma_c={1+i\over 2}$ is also needed in the
text. The following properties can be proven using the relation between
$\vartheta$-functions and elliptic integrals given below (A.16),
$$\vartheta_3^4(\sigma_c)=-\vartheta_4^4(\sigma_c)=
{i\over 4\pi^3}\Bigl\{\Gamma\Bigl({1\over 4}\Bigr)\Bigr\}^4,\autoeqA{A} $$
\newcount\Gammatheta\Gammatheta=\equnoA
(where $\Gamma\left({1\over 4}\right)\approx 3.626$ 
is the usual Gamma function)
from which we can conclude, using (A.\the\thetarelation), that
$$\vartheta_2^4(\sigma_c)=2\vartheta_3^4(\sigma_c)\quad\Rightarrow\quad f(\sigma_c)={1\over 4}.
\autoeqA{A} $$
\newcount\quarterf\quarterf=\equnoA 
With the help of (A.\the\fprime) above, a Taylor expansion for $f(\sigma)$ around
$\sigma_c={1+i\over 2}$ can now be developed, using (A.\the\thetadiff), 
(A.\the\Gammatheta) and  (A.\the\quarterf),
$$\eqalign{
f(\sigma_c + \varepsilon)&={1\over 4}\left\{1-2i\pi\;\vartheta_3^4(\sigma_c)
\left({\vartheta_3^\prime\over\vartheta_3}
   -{\vartheta_4^\prime\over\vartheta_4}\right)\Vline\sub{\sigma_c}\varepsilon^2 +\cdots\right\} \cr
&={1\over 4}\left\{1+{\pi^2\over 2}\vartheta_3^4(\sigma_c)
\vartheta_2^4(\sigma_c)\;\varepsilon^2+\cdots\right\}\cr
&={1\over 4}-{1\over 64\pi^4}
\left\{\Gamma\left({1\over 4}\right)\right\}^8\varepsilon^2+\cdots\; .\cr } \autoeqA{A}$$
\newcount\ftaylor\ftaylor=\equnoA
\smallskip
The Jacobi $\vartheta$-functions are related to complete elliptic integrals of
the second kind,
$$K(k)=\int_0^{\pi/2}{d\varphi\over\sqrt{1-k^2\sin^2\varphi}},\autoeqA{A} $$
by the following formula, [\the\WW] p.479,
$$\vartheta_2=\sqrt{2kK(k)\over\pi},\qquad
\vartheta_3=\sqrt{2K(k)\over\pi},\qquad
\vartheta_4=\sqrt{2k^\prime K(k)\over\pi},\autoeqA{A}$$
\newcount\thetaKA\thetaKA=\equnoA
where the modulus $k$ is related to $\sigma$ by $\e^{-\pi{K^\prime\over K}}=\e^{i\pi\sigma}$
and $K^\prime(k)=K(k^\prime)$, with $(k^\prime)^2:=1-k^2$ the complementary modulus.
\smallskip
The elliptic integral $K(k)$ has the following expansions, 
\autoref\newcount\GradRyz\GradRyz=\refno (8.113.1) and (8.113.3),
$$\eqalign{
K(k)&={\pi\over 2}\left(1+{1\over 4}k^2+\cdots\right)\hskip 4.3cm\vert k\vert<<1,\cr
K(k)&=\ln\left(4\over k^\prime\right)
+{1\over 4}\left\{ \ln\left( {4\over k^\prime} \right)-1\right\}(k^\prime)^2+\cdots
\qquad\qquad \vert k^\prime\vert<<1.\cr}\autoeqA{A} $$
\newcount\Kexp\Kexp=\equnoA
Thus
$$K^\prime(k)=\ln\left(4\over k\right)
+{1\over 4}\left\{ \ln\left( {4\over k} \right)-1\right\}k^2+\cdots
\qquad\qquad \vert k\vert<<1.\autoeqA{A} $$
\newcount\Kpexp\Kpexp=\equnoA
In addition the following relations are needed in the text,
$$K\left({1\over k}\right)=k\{K(k)+iK^\prime(k)\}, \autoeqA{A} $$
\newcount\Kinv\Kinv=\equnoA
$$K\left(i{k\over k^\prime}\right)=k^\prime K(k), \autoeqA{A} $$
\newcount\Kim\Kim=\equnoA
[\the\GradRyz] (8.128.1) and (8.128.3) (beware of the misprint in equation (8.128.3)
of the fourth edition).
\smallskip
Finally, using (8.129.1) of [\the\GradRyz],
$$K\left({1\over\sqrt{2}}\right)
={1\over 4\sqrt\pi}\left\{\Gamma\left({1\over 4}\right)\right\}^2 \autoeqA{A} $$
and (A.\the\Kinv) with (A.\the\thetaKA) reproduces equation (A.\the\Gammatheta) above,
since $\sigma={1+i\over 2}$ corresponds to $k=\sqrt{2}$.
\bigskip
\vfill\eject
{\bf References}
\bigskip
\item{[\the\LutkenRossa]} C.A.~L\"utken and G.G.~Ross, Phys. Rev. {\bf B45}, 11837
(1992)
\smallskip
\item{[\the\LutkenRossb]} C.A.~L\"utken and G.G.~Ross, Phys. Rev. {\bf B48}, 2500
(1993)
\smallskip
\item{[\the\WilcekZee]} A.~Shapere and F.~Wilczek, Nuc. Phys {\bf B320}, 669 (1989)
\smallskip
\item{[\the\GeorglinWallet]} Y.~Georgelin and J-C.~Wallet, Phys. Lett. {\bf A224}, 303
(1997);
Y.~Georgelin, T.~Masson and J-C.~Wallet, J. Phys. {\bf A30}, 5065 (1997)
\smallskip
\item{[\the\Lutken]} C.A.~L\"utken, Nuc. Phys. {\bf B396}, 670 (1993)
\smallskip
\item{[\the\BD]} B.P.~Dolan, {\sl Duality and the Modular Group in the Quantum Hall Effect}
\hfill\break
(cond-mat/9805171)
\smallskip
\item{[\the\Ruzin]}  A.M.~Dykhne and I.M.~Ruzin, Phys. Rev. {\bf B50}, 2369 (1994);\hfill\break
I.~Ruzin and S.~Feng, Phys. Rev. Lett. {\bf 74}, 154 (1995)
\smallskip
\item{[\the\KLZ]} S.~Kivelson, D-H.~Lee and S-C.~Zhang, Phys. Rev. {\bf B46},
2223 (1992)
\smallskip
\item{[\the\LutkenBurgess]} C.P.~Burgess and C.A.~L\"utken, Nuc. Phys. {\bf B500}, 367 (1997) 
\smallskip
\item{[\the\Ritz]} A.~Ritz, {\sl On the Beta-Function in N=2 Supersymmetric Yang-Mills
Theory}\hfill\break 
Phys. Lett. {\bf B434} 54, (1998) (hep-th/9710112)
\smallskip
\item{[\the\LutkenLatorre]} J.I.~Latorre and  C.A.~L\"utken, Phys.Lett. {\bf B421}, 217 (1998)  
(hep-th/9711150)
\smallskip
\item{[\the\Shaharetal]} D.~Shahar, D.~C.~Tsui, M.~Shayegan, E.~Shimshoni and
S.~L.~Sondhi, Phys. Rev. Lett. {\bf 79}, 479 (1997) (cond-mat/9611011)
\smallskip
\item{[\the\Pruiskena]} A.M.M.~Pruisken, Phys. Rev. Lett. {\bf 61}, 1297 (1988)
\smallskip
\item{[\the\STSCSS]} D.~Shahar, D.C.~Tsui, M.~Shayegan, J.E.~Cunningham,
E.~Shimshoni and S.L.~Sondhi, Solid State Comm. {\bf 102} (1997) 817  (cond-mat/9607127);
\hfill\break
D.~Shahar, D.C.~Tsui, M.~Shayegan, E.~Shimshoni and S.L.~Sondhi,  Science {\bf 274}, (1996) 589
(cond-mat/9510113)
\smallskip
\item{[\the\Khem]} D.E.~Khmel'nitskii, Pis'ma Zh. Eksp. Teor. Fiz {\bf 38}, 454
(1983) \hfill\break
(JETP Lett. {\bf 38}, 552 (1983)
\smallskip
\item{[\the\QPT]} S.L.~Sondhi, S.M.~Girvin, J.P.~Carini and D.~Shahar, Rev. Mod. Phys.
{\bf 69}, 315 (1997)
\smallskip
\item{[\the\Pruiskenb]} {\sl The Quantum Hall Effect}, Eds. R.E.~Prange and S.M.~Girvin, Springer, (1987)
\smallskip
\item{[\the\SW]} N.~Seiberg and E.~Witten, Nuc. Phys. {\bf B426}, 19 (1994) (hep-th/9407087)
\smallskip
\item{[\the\MontonenOlive]} C.~Montonen and D.~Olive, Phys. Lett. {\bf B72}, 117 (1997)
\smallskip
\item{[\the\CardyRabinovici]} J.L.~Cardy and E.~Rabinovici, Nuc. Phys. {\bf B205}, 1
(1982); J. L. Cardy, Nuc. Phys. {\bf B205}, 17 (1982)
\smallskip
\item{[\the\Rankin]} R.A.~Rankin, {\sl Modular Forms and Functions}, C.U.P.
(1977)
\smallskip
\item{[\the\Damgaard]} P.H.~Damgaard and P.E.~Haagensen, J. Phys. {\bf A30}, 4681 (1997)\hfill\break (cond-mat/9609242)
\smallskip
\item{[\the\YSBTS]} K.~Yang, D.~Shahar, R.N.~Bhatt, D.C.~Tsui and M.~Shayegan,
cond-mat/9805341
\smallskip
\item{[\the\HHB]} Y.~Huo. R.E.~Hetzel and R.N.Bhatt, Phys. Rev. Lett. {\bf 70}, 481 (1993)
\smallskip
\item{[\the\WW]} E.T.~Whittaker and G.N.~Watson, {\sl A Course of Modern Analysis}, C.U.P.
(1940)
\smallskip
%\item{[\the\PrangeGirvin]} R.E.~Prange and S.M.~Girvin (ed.), 
%{\sl The Quantum Hall Effect}, (1987) Springer 
%\smallskip
\item{[\the\SHLTSR]} D.~Shahar, M.~Hilke, C.C.~Li, D.C.~Tsui, S.L.~Sondhi and M.~Razeghi,\hfill\break
cond-mat/9706045
\smallskip
\item{[\the\HSSTXM]} M.~Hilke, D.~Shahar, S.H.~Song, D.C.~Tsui, Y.H.~Xie and D.~Monroe,\hfill\break
cond-mat/9708239
\smallskip
\item{[\the\Baletal]} S.C.~Zhang, Int. J. Mod. Phys. {\bf B6}, 25 (1992)
%A.P.~Balachandran, L.~Chandar and B.~Sathiapalan, Nuc. Phys. {\bf 443 [FS]}, 465, (1995)  
\smallskip
\item{[\the\GradRyz]} I.S.~Gradshteyn and I.M.~Ryzhik, {\sl Tables of Integrals, Series and Products},
5th edition, Academic Press (1994)

\vfill\eject
\baselineskip = 0.67\baselineskip
\includegraphics{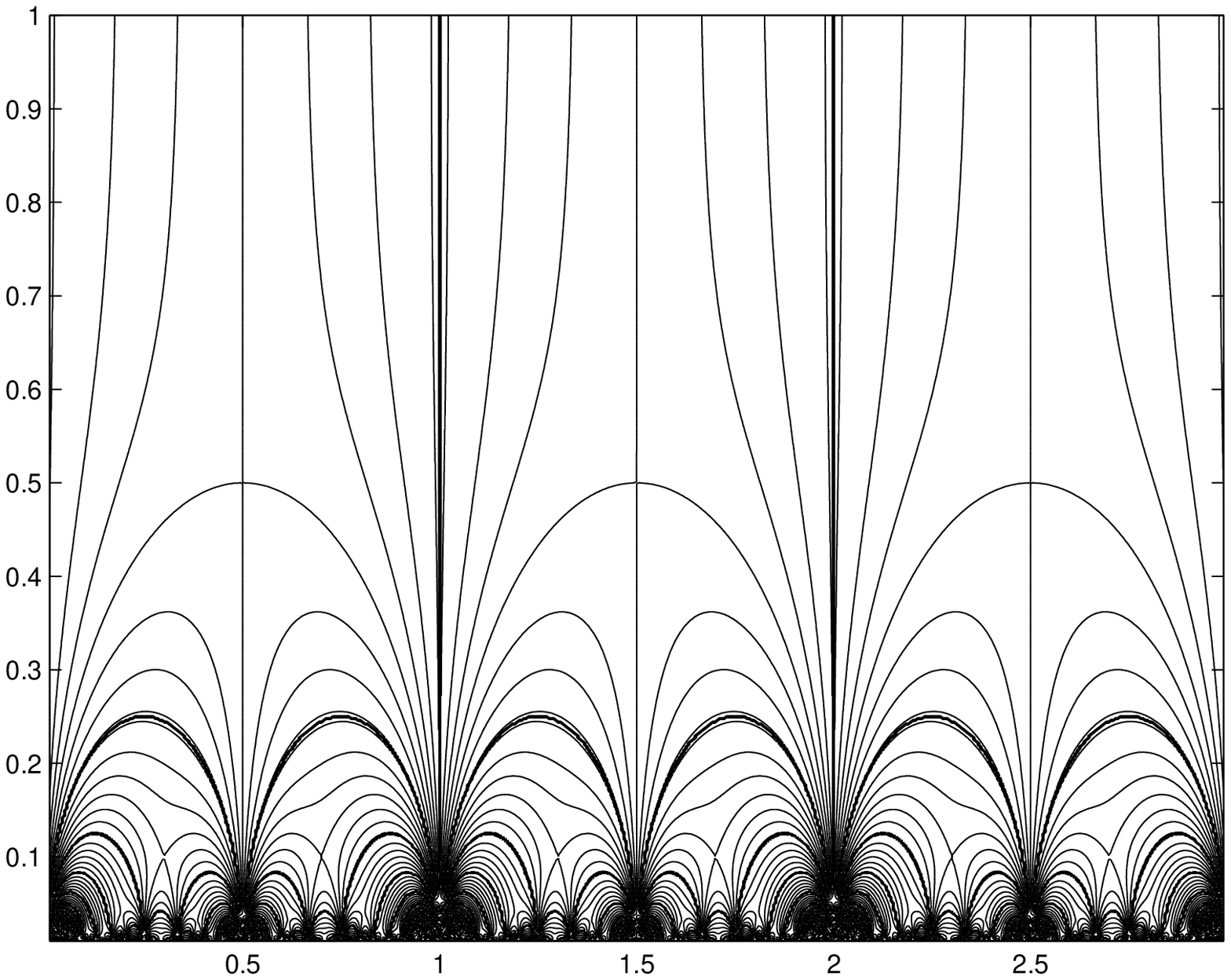}
\vskip 1cm\centerline{}
\vskip 1cm \hskip .1cm$\sigma_{xx}$
\vskip 5cm
\hskip 5cm $\sigma_{xy}$
\vskip 1cm
\leftline{Figure 1: RG flow in the complex conductivity plane for the quantum Hall effect.}
\leftline{The thicker lines are cuts where the phase of $f(\sigma)$
jumps form $-\pi$ to $\pi$.}
\vfill\eject
\includegraphics{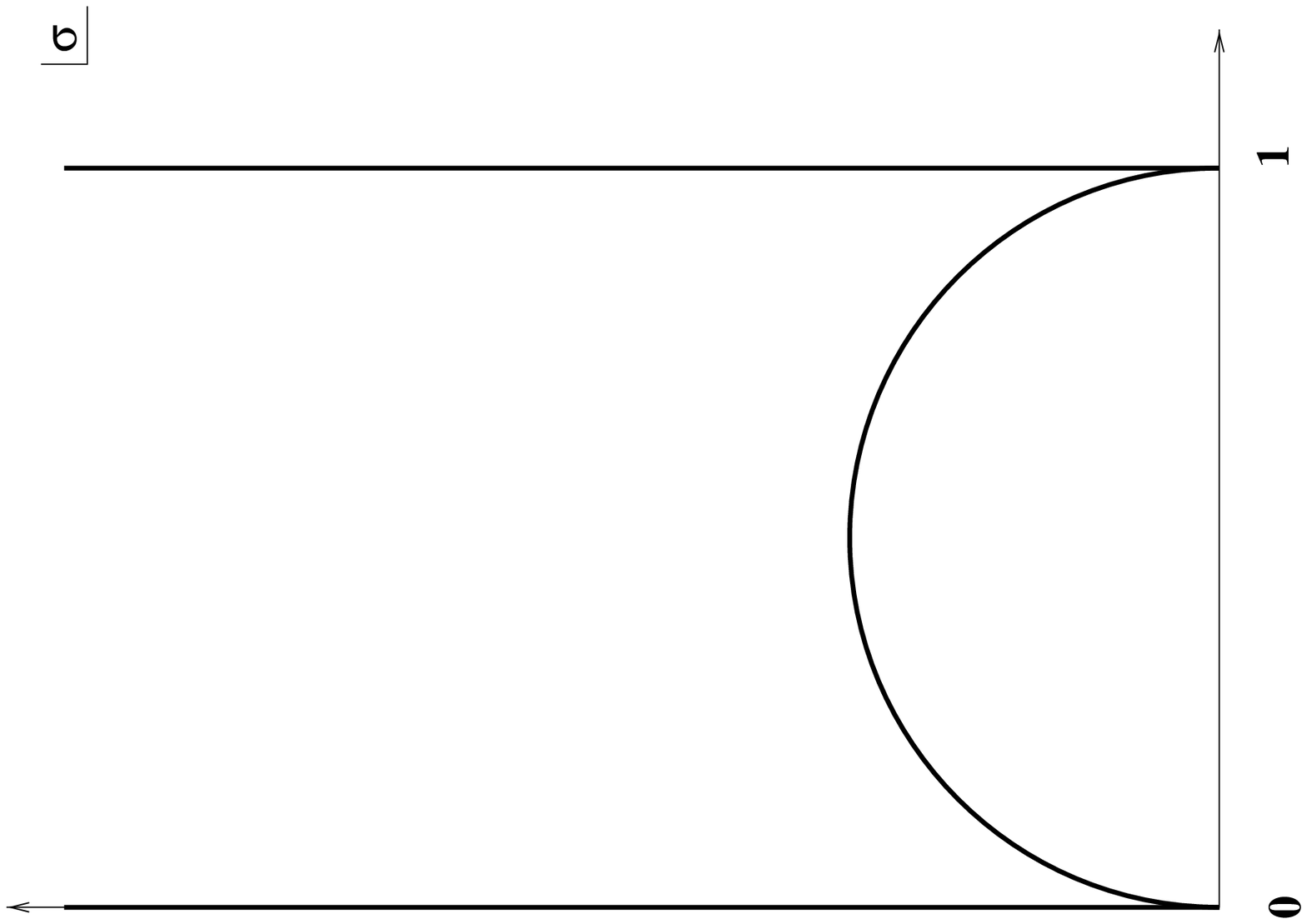}
\centerline{}
\vskip 10cm
\leftline{Figure 2: The fundamental domain for $\Gu$ is the vertical
strip of unit width}
\leftline{above the semi-circle of unit diameter centered at $\sigma=1/2$}
\vfill\eject

\includegraphics{fig3.ps}
\centerline{}
\vskip 2cm
\hskip .5cm$\rho_{xx}$
\vskip 3.7cm
\hskip 4.5cm $\Delta \nu$
\vskip 0.5cm
\leftline{Figure 3: Crossover of longitudinal resistivity for the insulator-QH transition}
\leftline{ $\nu:0\rightarrow 1$ at the four temperatures $T=42$, $84$,
$106$ and $145mK$ with $A=60$ and $\mu=0.50$.}
\leftline{To be compared with the experimental data in figure 2b of [\the\Shaharetal].} 
\vskip -8cm \hskip 7cm $\nu: 0\rightarrow 1$
\vfill\eject
\includegraphics{fig4a.ps}
\centerline{}
\vskip 2cm
\hskip .5cm$\rho$
\vskip 3.7cm
\hskip 5cm $\Delta \nu$
\vskip 0.5cm
\hskip 3.5cm{ }
\hskip -8cm
\vskip -7cm
\includegraphics{fig4b.ps}
\centerline{}
\vskip 10cm
\hskip .5cm$\sigma$
\vskip 3.7cm
\hskip 5cm $\Delta \nu$
\vskip 0.5cm
\leftline{Figure 4: Crossover of conductivity and resistivity for $\nu:1\rightarrow 2$}
\leftline{at the four temperatures $T=42$, $70$, $101$ and $137mK$ 
with $A=40$ and $\mu=0.50$.}
\leftline{To be compared with the experimental data in figure 1 of [\the\Shaharetal].} 
\vskip -16cm \hskip 7cm $\nu: 1\rightarrow 2$ 

\bye